\documentclass{ptapap}
\usepackage{color}

\usepackage{soul}

\author{Karolina Bąkowska}[CAMK]
\author{Thomas R. Marsh}[Warwick]
\affil[CAMK]{Nicolaus Copernicus Astronomical Center, 
  Bartycka 18, 00--716 Warszawa, Poland}
\affil[Warwick]{Department of Physics, University of Warwick, Coventry CV4 7AL, United Kingdom}

\title{Spectroscopy of the cataclysmic variable ES Ceti}

\begin{document}

\maketitle

\begin{abstract}

We present results of our spectroscopic campaign dedicated to the ultracompact binary ES Ceti. On the nights 2002 Oct. 27-28, 528 spectra were taken with the 6.5-meter telescope in Las Campanas Observatory in Chile. The averaged spectrum shown the double-peaked helium emission lines which imply the presence of an accretion disk in this system.

\end{abstract}

\section{Introduction}

Ultracompact binaries consist of a primary white dwarf with a companion component also at least partially degenerated. It is suggested that close double-degenerate binaries are possible progenitor populations of Type Ia supernovae and among detectable sources of gravitational wave radiation. 

The AM CVn stars are helium-rich binaries with their orbital periods ranging from 5 to 65 min. Almost all of them do not show traces of hydrogen in their spectra. They evolve through one or two common envelope (CE) events. Therefore, the AM CVn family is important for our understanding  of the  binary formation and the CE phase \citep[for review see:][]{2010PASP..122.1133S}.

Among the AM CVn stars there are two systems with the shortest known orbital periods: HM Cnc  \citep[324\,s,][]{2010ApJ...711L.138R} and V407 Vul \citep[569s,][]{1995A&A...297L..37H}. The exact nature of these 
objects is unknown. ES Ceti shows the 620\,s orbital period, which is only 51\,s greater than the one of V407 Vul. Hence, it is the perfect subject for a timing study based on photometric observations \citep{2011MNRAS.413.3068C} and preliminary analysis of the 
disk structure (this work).

\section{Averaged and Trailed Spectra}
\begin{figure}
\centering
   \includegraphics[width=0.75\textwidth]{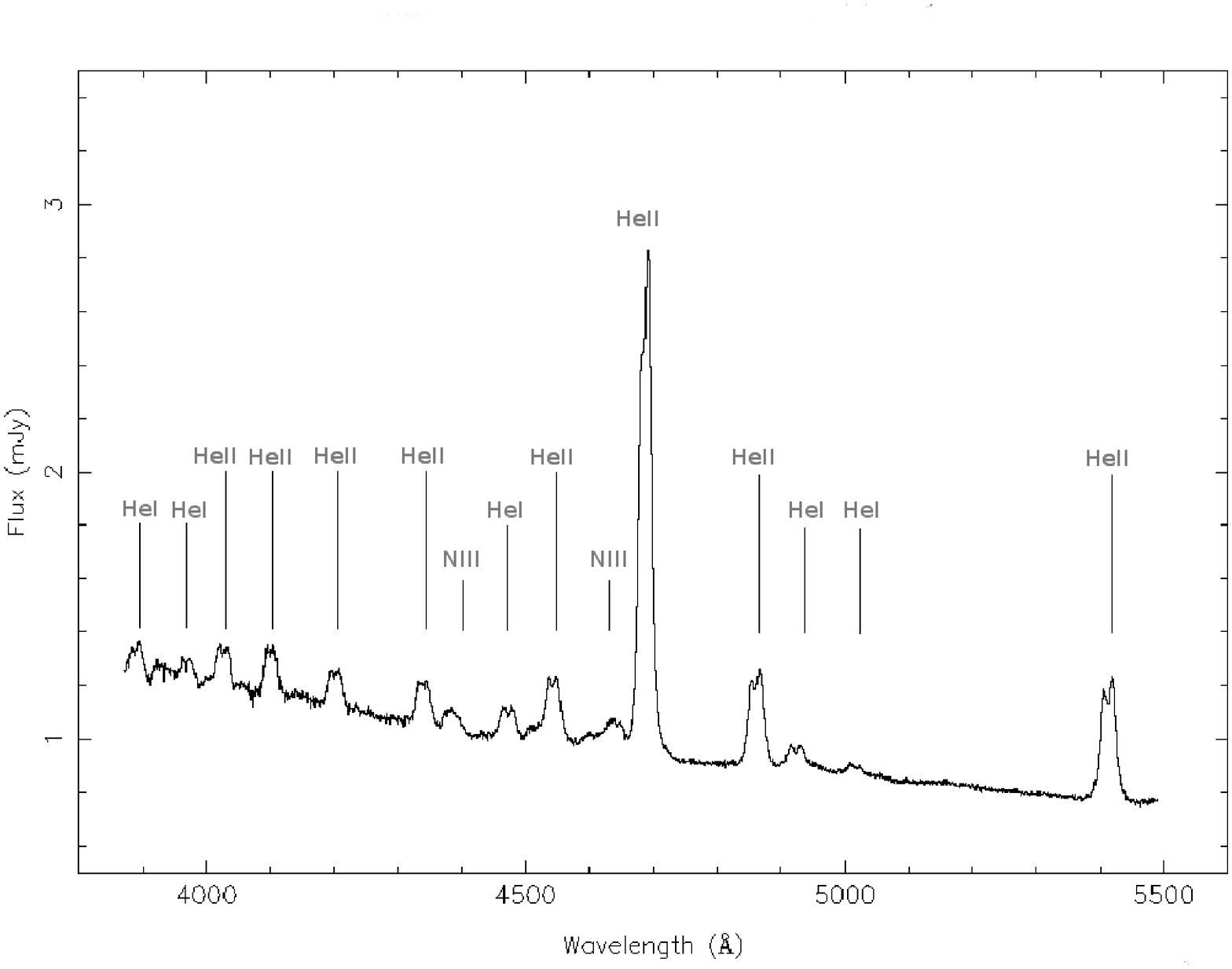}
   \caption{The mean spectrum of ES Ceti taken with Magellan on 27-28 Oct. 2002. All lines can be identified with He\,II, He\,I or N\,III.}
   \label{fig:ESCet_Average}
\end{figure}

The averaged Magellan spectrum of ES Ceti is shown in Fig.\ref{fig:ESCet_Average}. The spectrum is dominated by ionized helium emission lines. Lines of neutral helium and nitrogen N\,III are also present. 

\begin{figure}
\centering
\begin{minipage}{0.48\textwidth}
\includegraphics[width=\textwidth]{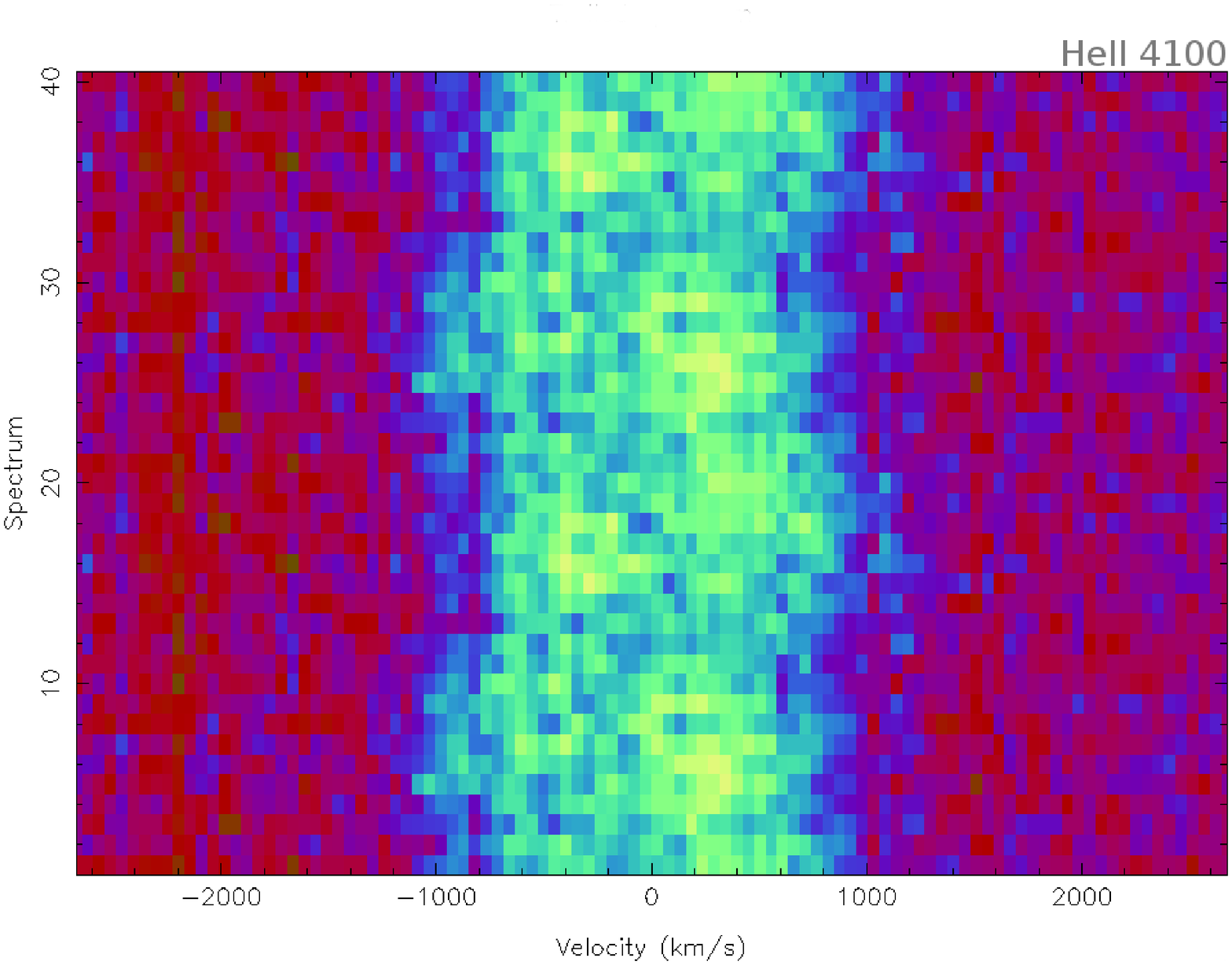}
\end{minipage}
\quad
\begin{minipage}{0.48\textwidth}
\includegraphics[width=\textwidth]{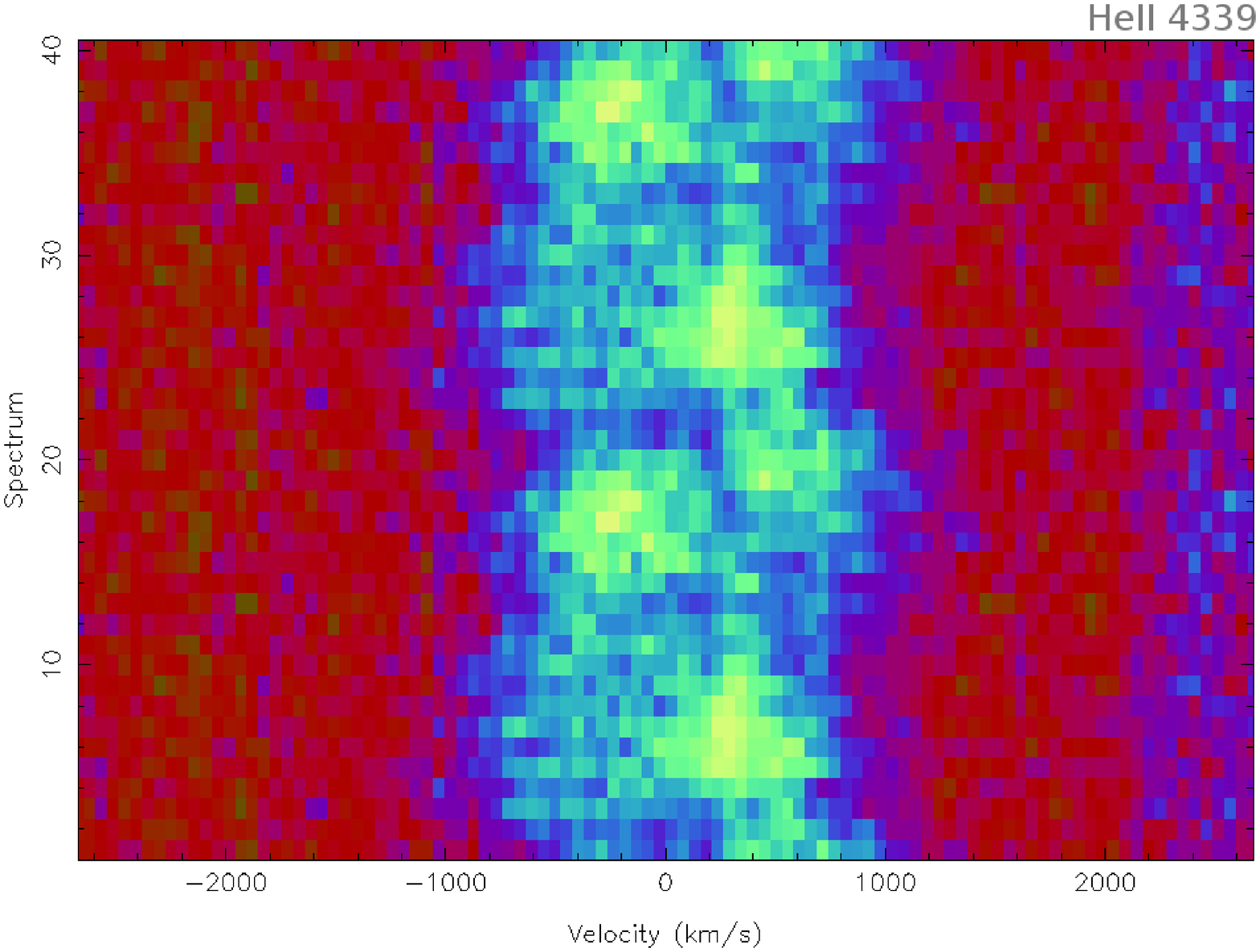}
\end{minipage}
\begin{minipage}{0.48\textwidth}
\includegraphics[width=\textwidth]{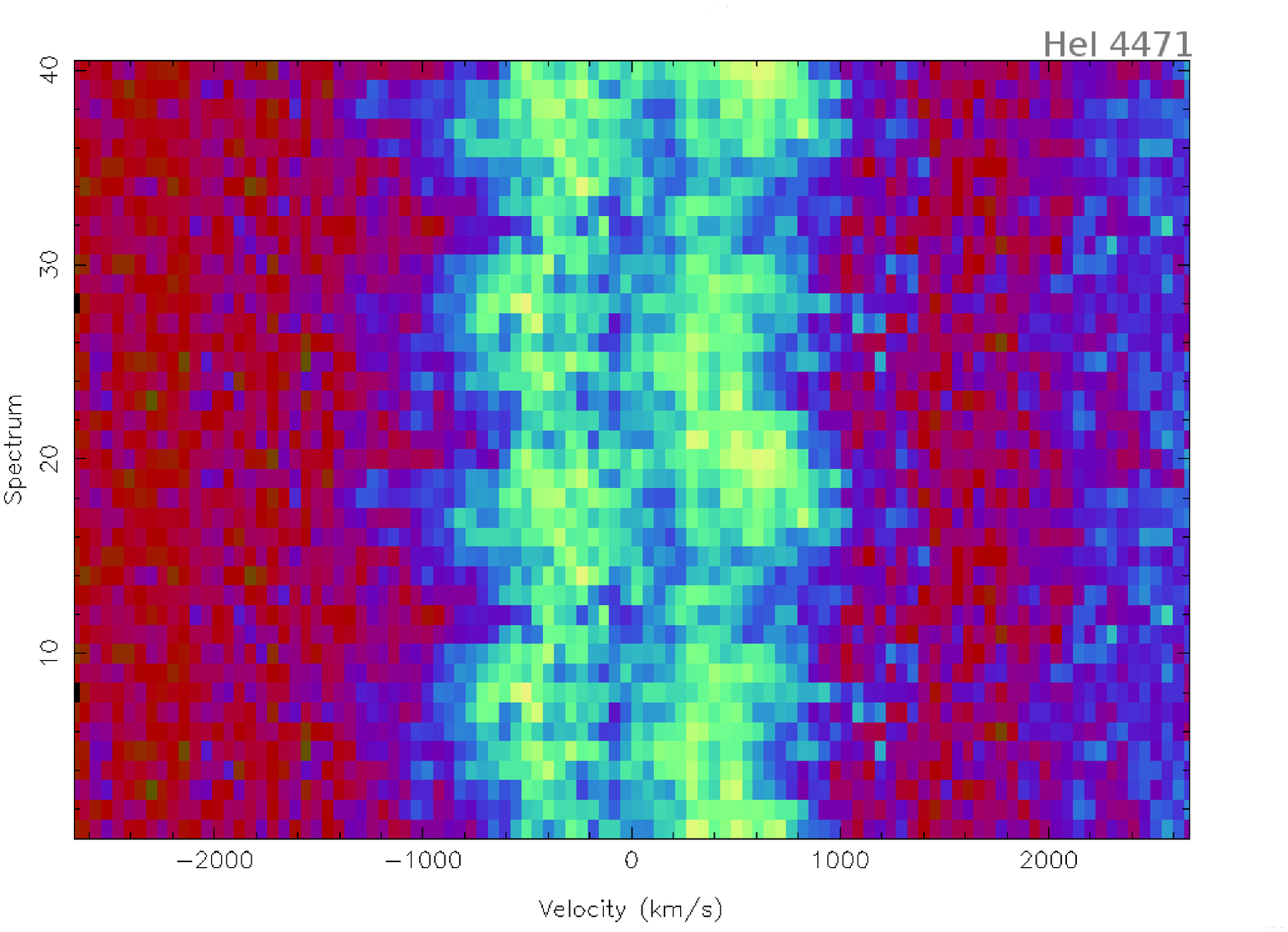}
\end{minipage}
\quad
\begin{minipage}{0.48\textwidth}
\includegraphics[width=\textwidth]{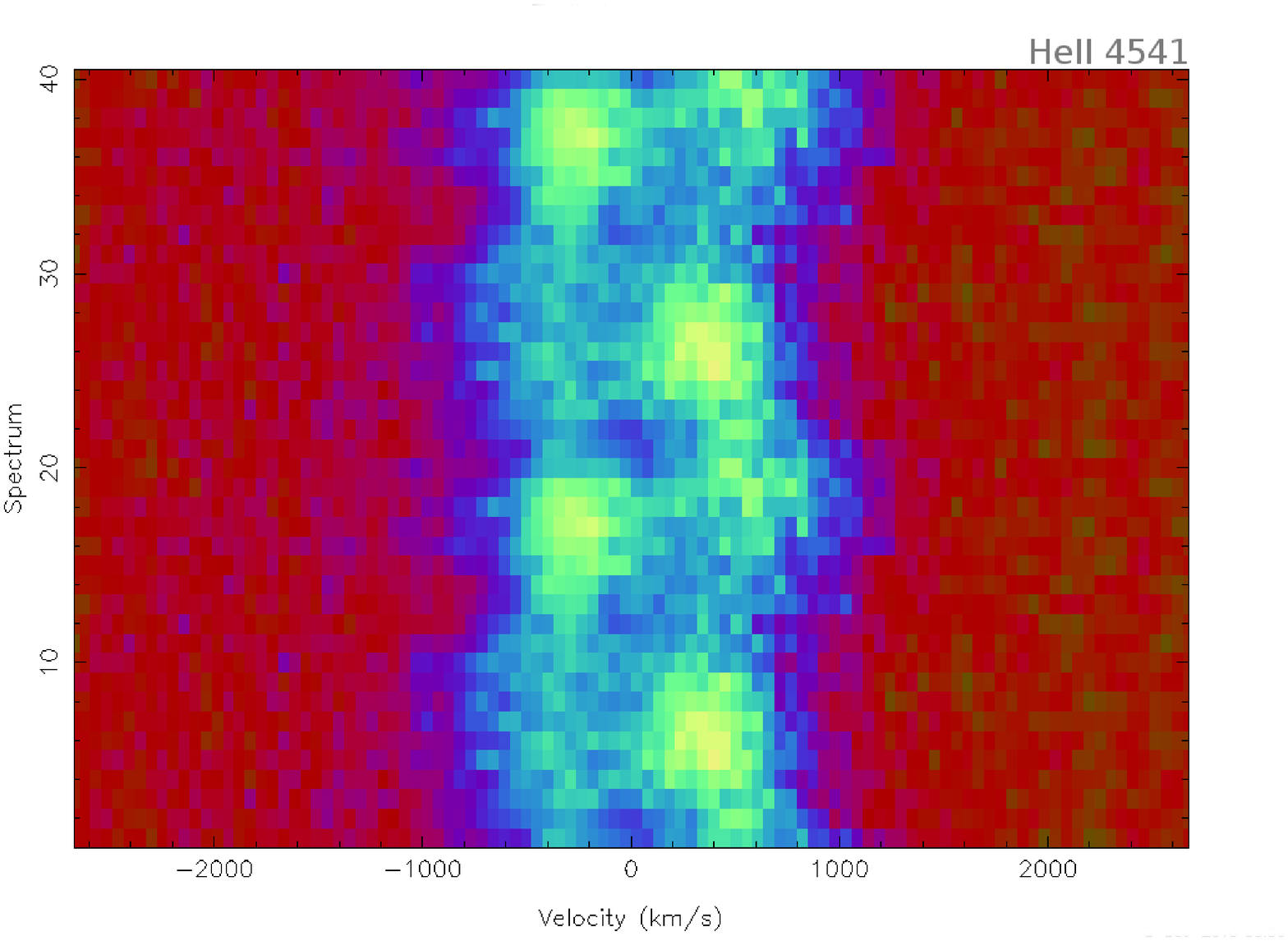}
\end{minipage}
\begin{minipage}{0.48\textwidth}
\includegraphics[width=\textwidth]{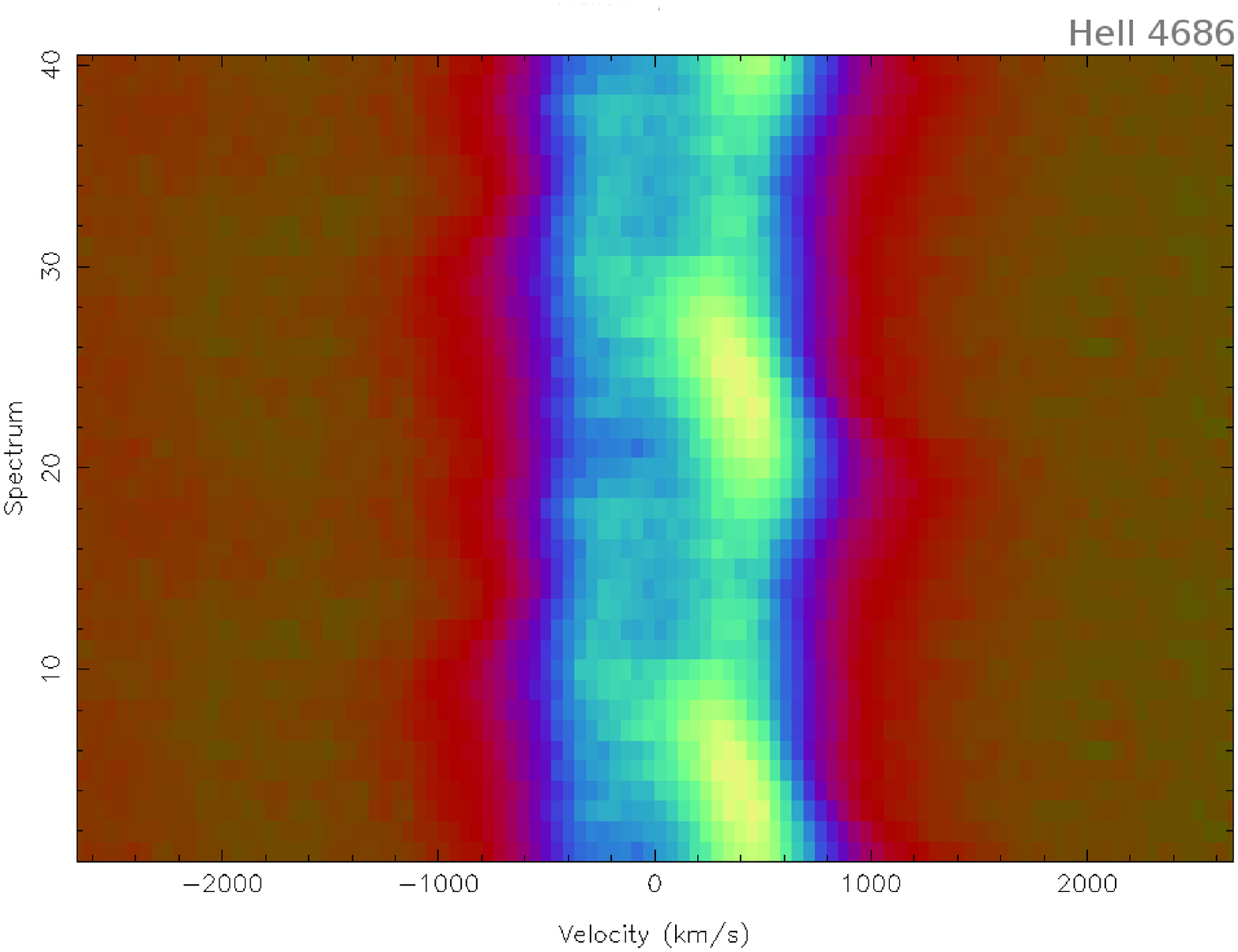}
\end{minipage}
\quad
\begin{minipage}{0.48\textwidth}
\includegraphics[width=\textwidth]{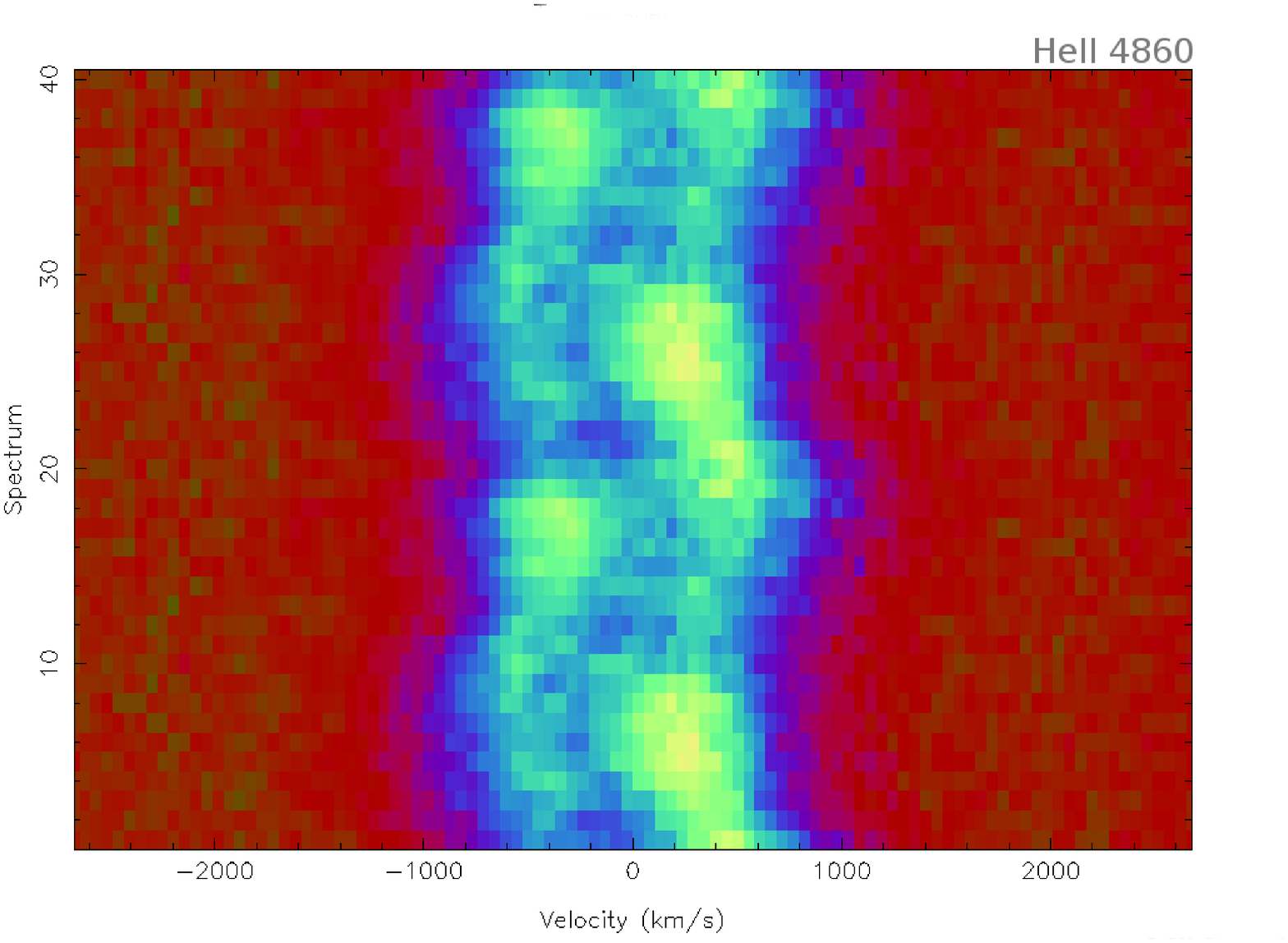}
\end{minipage}
\begin{minipage}{0.48\textwidth}
\includegraphics[width=\textwidth]{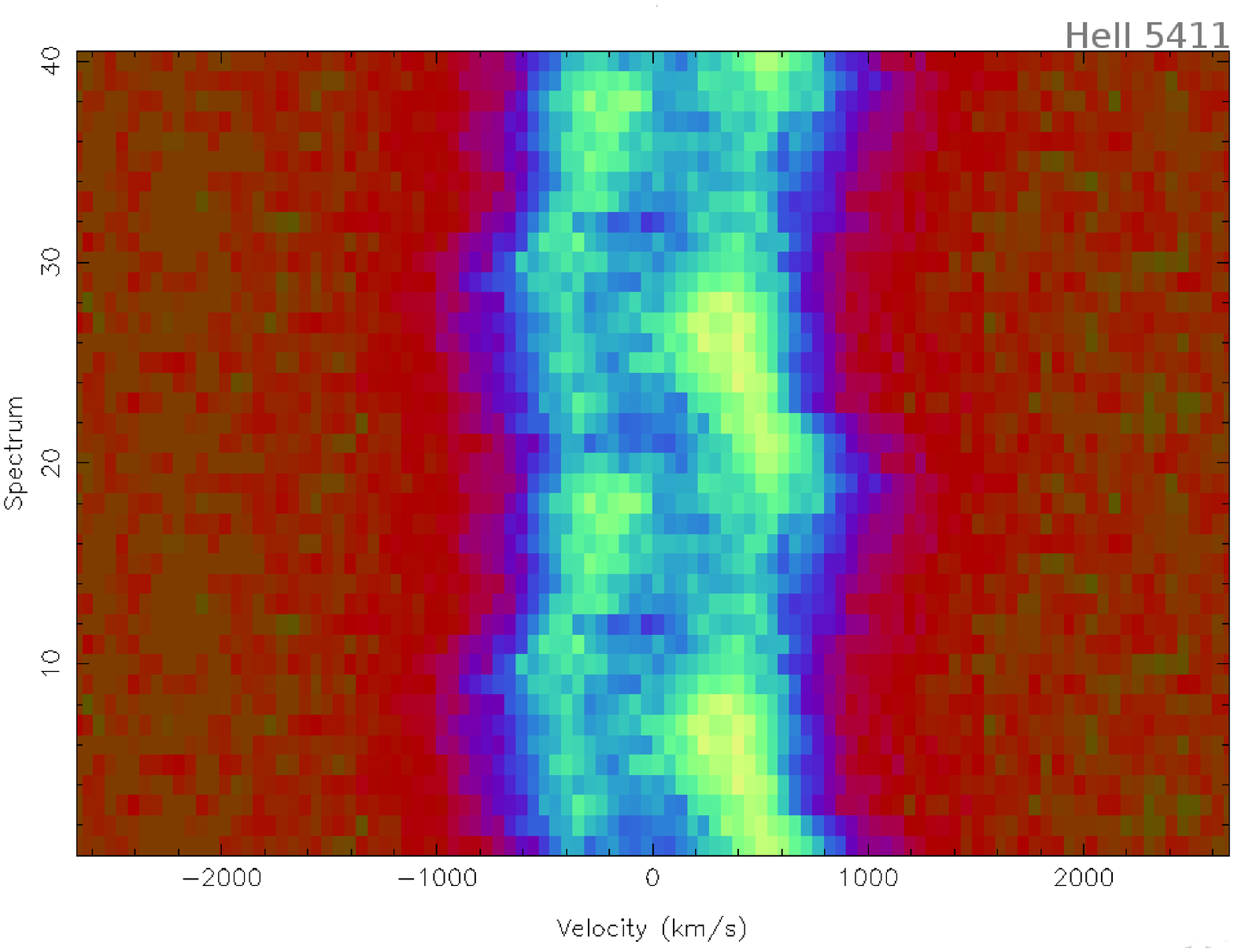}
\end{minipage}
\caption{The phase-folded, continuum-normalised and subtracted trailed spectra of ES Ceti.}
\label{fig:ESCet_Trail}
\end{figure}

Panels of Fig.\ref{fig:ESCet_Trail} show the time-resolved (trailed spectra) of ES Ceti, folded on the ephemeris given by \cite{2011MNRAS.413.3068C}. These data are the first spectroscopic observations confirming that the observed period is indeed orbital. However, the variations seen in trailed spectra are not classical, they are very clear and "S-wave"-like. Worth noting is their double-peaked structure, e.g. in the He\,II 5411 line.

\section{Summary and Future Work}

Based on the presented spectroscopic data, we concluded that the accretion in ES Ceti is via a disk, and we excluded the direct-impact scenario proposed by \cite{2005PASP..117..189E}.  Probably, the hot gas produces a "disk-like" signature similar to the one observed in HM Cnc system \citep{2010ApJ...711L.138R}.

The strong emission lines are among of the hallmarks of an accreting binaries. Hence, based on the trailed spectra of ES Ceti, we plan to create the equivalent Doppler maps. The method of Doppler tomography \citep[see:][]{1988MNRAS.235..269M} is a perfect tool which allows to track asymmetric structures in accretion disks and reveals details of the gas flow in a variety of systems. 

\acknowledgements{Project was supported by  Polish National Science Center grants awarded by decisions: DEC-2015/16/T/ST9/00174 and DEC-2015/18/A/ST9/00578 for KB.}

\bibliographystyle{ptapap}
\bibliography{ptapapdoc}

\end{document}